\documentclass[prl,reprint,showpacs,floatfix,superscriptaddress]{revtex4-1}
\usepackage[pdftex]{graphicx} 
\usepackage{dcolumn} 
\usepackage{bm} 
\usepackage{amssymb} 
\usepackage{amsmath} 
\usepackage{footmisc}
\usepackage[latin1]{inputenc}
\usepackage{tikz}
\usetikzlibrary{shapes,arrows}

\usepackage{enumitem}

\begin{document}
\title{Accuracy of electron densities obtained via Koopmans-compliant hybrid functionals}
\date{\today}
\author{A.\ R.\ Elmaslmane}
\thanks{These authors contributed equally to this work}
\affiliation{Department of Physics, University of York, Heslington, York YO10 5DD, United Kingdom}
\author{J.\ Wetherell}
\thanks{These authors contributed equally to this work}
\affiliation{Department of Physics, University of York, Heslington, York YO10 5DD, United Kingdom}
\affiliation{European Theoretical Spectroscopy Facility}
\author{M.\ J.\ P.\ Hodgson}
\affiliation{Max-Planck-Institut f\"ur Mikrostrukturphysik, Weinberg 2, D-06120 Halle, Germany}
\affiliation{European Theoretical Spectroscopy Facility}
\author{K. P.\ McKenna}
\affiliation{Department of Physics, University of York, Heslington, York YO10 5DD, United Kingdom}
\author{R.\ W.\ Godby}
\affiliation{Department of Physics, University of York, Heslington, York YO10 5DD, United Kingdom}
\affiliation{European Theoretical Spectroscopy Facility}

\begin{abstract}
We evaluate the accuracy of electron densities and quasiparticle energy gaps given by hybrid functionals by directly comparing these to the exact quantities obtained from solving the many-electron Schr\"odinger equation. We determine the admixture of Hartree-Fock exchange to approximate exchange-correlation in our hybrid functional via one of several physically justified constraints, including the generalized Koopmans' theorem. We find that hybrid functionals yield strikingly accurate electron densities and gaps in both exchange-dominated and correlated systems. We also discuss the role of the screened Fock operator in the success of hybrid functionals. 
\end{abstract}
\maketitle

A key measure of success for any electronic-structure theory is 
its ability to yield accurate electron densities and energies for many-electron systems.
For example, Kohn-Sham (KS) density functional theory (DFT) \cite{PhysRev.136.B864,PhysRev.140.A1133} 
is in principle exact, but the use of an approximate exchange-correlation (xc) potential, such as the local density approximation (LDA) \cite{PhysRevA.38.3098} or the generalized gradient approximation (GGA) \cite{PhysRevLett.77.3865}, is associated with a self-interaction error which can cause the spurious delocalization of localized charge \cite{PhysRevB.23.5048} and incorrect dissociation behavior for molecules \cite{PhysRevLett.87.133004}. Recently, hybrid functionals that mix Hartree-Fock (HF) exchange with a (semi-)local approximation (such as the LDA or GGA) \cite{doi:10.1063/1.472933} have become popular as an alternative approach to xc. However, hybrids introduce at least one additional parameter, the mixing parameter $\alpha$. This is often determined empirically, e.g., via experimental data, or through the adiabatic connection \cite{doi:10.1063/1.464304}. We determine $\alpha$ using a group of more physically justified constraints, including the generalized Koopmans' theorem \cite{doi:10.1021/ct2009363,PhysRevB.94.035140,PhysRevB.88.081204,PhysRevB.80.085202,PhysRevB.82.115121}. While it has been shown that this constrained hybrid approach results in ionization energies and band gaps close to experimental values \cite{doi:10.1021/ct2009363,PhysRevB.88.081204}, to date the electron density of this approach has not been directly compared to the exact density. 

As Medvedev {\it et al.} \cite{Medvedev49} argue, progress in the accuracy of electronic structure calculations requires improvements in both energies \textit{and} densities. Srebro {\it et al.} indirectly assessed densities obtained via hybrid functionals using the electric field gradient at the nucleus \cite{doi:10.1021/ct200764g}. Reference \onlinecite{doi:10.1063/1.2821123} obtained densities from popular empirical hybrid functional parameterizations and found sensitivity to the value of the various mixing parameters. Good agreement between hybrid and CCSD densities has been found for the CO molecule \cite{doi:10.1080/00268970009483369}.

In order to address the density more directly, we consider a set of model systems where the many-body problem can be solved exactly for a small number of electrons, allowing for a direct comparison of densities, energy gaps and ionization potentials (IPs) obtained from the constrained hybrid approach to the exact values.  We show that an \textit{ab initio} determination of $\alpha$ results in hybrid functionals yielding extremely accurate densities and gaps.

The exact total energy $E$ (of a many-electron system) is piecewise linear with respect to the number of electrons, $N$ \cite{PhysRevLett.49.1691,PhysRevA.30.2745}. In exact KS DFT, the slope of each straight-line segment $\partial E/\partial N$ is shown by Janak's theorem to equal the highest (partly) occupied molecular orbital (HOMO) eigenvalue \cite{PhysRevB.18.7165}. The usual approximate density functionals (LDA and GGAs), and HF, exhibit nonzero curvature $\partial^2 E/\partial N^2$, which can lead to qualitatively wrong physical behavior \cite{PhysRevLett.100.146401,Cohen792,doi:10.1021/cr200107z}. The curvatures are of opposite signs which means that hybrid approximations benefit from a partial cancellation of these errors \footnote{An illustration of the curvature is shown in supplemental material.} \cite{doi:10.1021/ct2009363,PhysRevB.94.035140}.

The exact total energy difference $E(N\!-\!1)-E(N)$ is both the ionization energy of the $N$--electron system, $I(N)$, and the electron affinity of the $(N\!-\!1)$--electron system, \mbox{$A(N\!-\!1)$}. 
In HF, the equivalent of Janak's theorem \cite{doi:10.1063/1.3702391} shows that the slope $(\partial E/\partial N)_{N-\delta}$ is equal to the HOMO eigenvalue, and $(\partial E/\partial N)_{N+\delta}$ to the LUMO eigenvalue.
In exact KS DFT, the LUMO eigenvalue differs from the negative electron affinity $-A$ by a discontinuity, $\Delta$, in the xc potential \cite{PhysRevLett.49.1691}. Thus all three quantities $\varepsilon_N(N\!-\!1)+\Delta$ \footnote{$\varepsilon_N(N\!-\!1)$ denotes the $N^\text{th}$ eigenvalue for the $(N\!-\!1)$--electron system.}, $\varepsilon_N(N)$ and $E(N)-E(N\!-\!1)$ should, in principle, be equal, where $\Delta$ is non-zero for exact DFT methods. But for approximate methods such as hybrids where exchange and correlation are explicitly analytical functionals of the single-particle orbitals and therefore exhibit zero derivative discontinuity $\Delta$, the first quantity becomes $\varepsilon_N(N\!-\!1)$ \cite{doi:10.1063/1.3702391, Perdew14032017, PhysRevB.53.3764}. We may therefore identify three requirements, 
\begin{enumerate}[label=(\Alph*)]\vspace{-0.075cm}
\item $\varepsilon_N(N\!-\!1) = -A(N\!-\!1)\equiv E(N)-E(N\!-\!1)$,
\item $\varepsilon_N(N\!-\!1)=\varepsilon_N(N)$,
\item $\varepsilon_N(N) = -I(N)\equiv E(N)-E(N\!-\!1)$,\vspace{-0.075cm}
\end{enumerate}
which may be used to constrain a hybrid calculation by enforcing internal consistency.
In practice, the parameter $\alpha$ of the basic hybrid approach provides a single degree of freedom and so can be used to impose (A) the LUMO-$A$ condition \textit{or} (B) the LUMO-HOMO condition \textit{or} (C) the HOMO-$I$ condition, or generalized Koopmans' theorem (GKT).
The merits, as regards electron energies, of satisfying the last two conditions using a more elaborate hybrid form has been investigated \cite{doi:10.1021/jz3015937,PhysRevB.84.075144,doi:10.1021/ja8087482}

A key point regarding the hybrid approach is that the derivative discontinuity $\Delta$ in the xc potential not only \textit{is} zero, but also \textit{should be} zero, when viewed from the perspective of many-body perturbation theory. This is most clearly seen by noting that the description of exchange and correlation in the hybrid approach includes a reduced-strength Fock operator, essentially mimicking the screened exchange operator that is at the heart of the well-known $GW$ approximation to the self-energy operator \cite{PhysRev.139.A796,1998RPPh...61..237A,PhysRevLett.105.266802,PhysRevB.88.165122}, plus LDA exchange and correlation reduced in strength. This identification of the hybrid approach's ``self-energy'' as a screened-exchange approximation to the exact self-energy $\Sigma_\mathrm{xc}$, as noted by other authors \cite{MarquesPhysRevB.83.035119,BrawandPhysRevX.6.041002}, means that $\Sigma_\mathrm{xc}$ would yield exact electron addition and removal energies through its one-electron eigenvalues that then acquire the significance of quasiparticle energies. Hence in both the $N$ and ($N\!-\!1$)--particle systems both the HOMO and LUMO energies may be regarded as fairly sophisticated approximations to the ionization potential and electron affinity, and therefore require no $\Delta$ correction.

The hybrid functional that we use for our main tests straightforwardly mixes HF with an LDA xc potential:
\begin{gather}\label{eqn:hybrid-potential}
V_\mathrm{xc}^\text{HYB}(\alpha)=\alpha V_\mathrm{x}^\text{HF} + (1-\alpha)V_\mathrm{xc}^\text{LDA},
\end{gather}
where $V_\mathrm{xc}^\text{HYB}$, $V_\mathrm{xc}^\text{LDA}$ and $V_\mathrm{x}^\text{HF}$ denote the hybrid and LDA xc potentials \footnote{The LDA used in this work is parameterized from finite slabs \cite{PhysRevB.94.205134}; our testing has shown these give indistinguishable results compared to homogeneous electron gas (HEG)-based LDAs. See supplemental material for further information.}  and the non-local HF exchange potential, respectively. This has the advantage of focusing more on the variational power of HF for exchange-dominated systems and accommodating better the cross-over between exchange and correlation when the LDA is applied to inhomogeneous systems. We also explore the retention of the full LDA correlation potential, mixing only the exchange terms, in common with other hybrid functionals such as PBE0 \cite{doi:10.1063/1.472933}. 

We assess hybrid functionals both in systems where correlation is relatively unimportant (``exchange-dominated'')  and systems in which correlation is more significant. The exact many-body wavefunction (used to compute the exact density) is obtained by direct solution of the many-body Schr\"odinger equation using the iDEA code \cite{PhysRevB.88.241102}. The electrons interact via the softened Coulomb interaction $(|x-x'|+1)^{-1}$ and are treated as spinless in order to model more closely the richness of correlation found in systems containing a large number of electrons.

\begin{figure}[t!]
\centering
\includegraphics[width=3.3in]{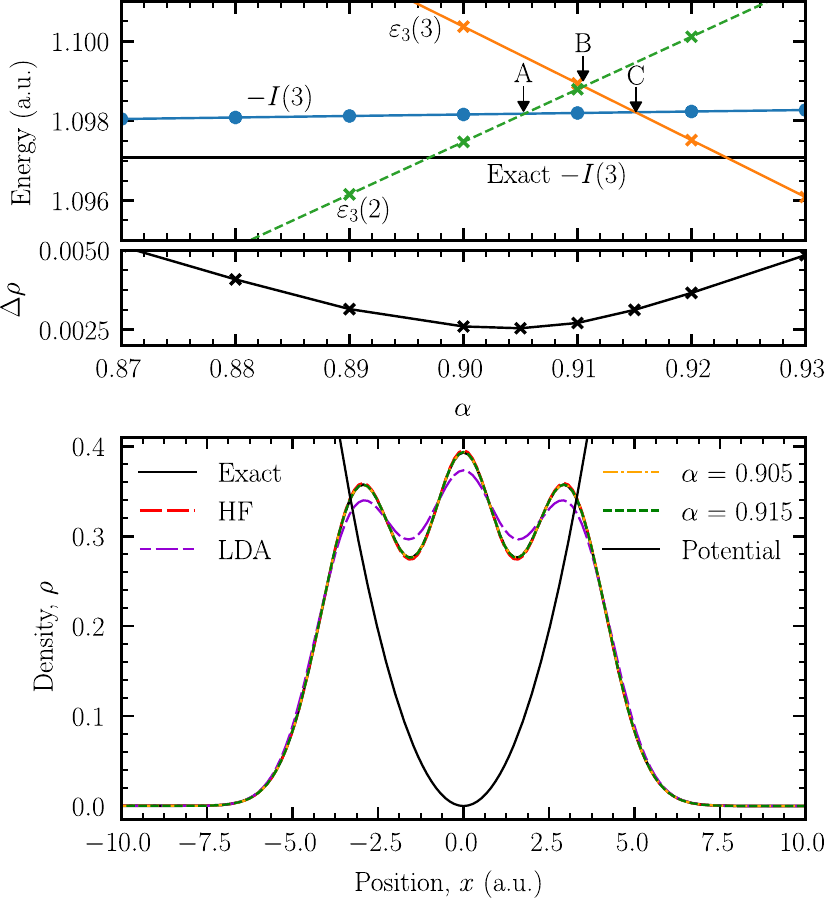}
\caption{\small (Upper) The variation in hybrid ionization energy $I(3)(=A(2))$, exact $I(3)(=A(2))$, $\varepsilon_3(3)$ and $\varepsilon_3(2)$ with $\alpha$ are illustrated for three electrons in an harmonic oscillator with $\omega=0.25$, an exchange-dominated system. Energies are in Hartree atomic units. There are three `crossing points': (A) $A$-LUMO, (B) HOMO-LUMO and (C) $I$-HOMO. (Center) The integrated absolute error in the density $\Delta\rho$ is shown for each value of $\alpha$. This is defined as $\int|\rho^{\text{EXT}}(x)-\rho^{\text{HYB}}(x)|dx$ where the $\rho^\text{EXT}$ and $\rho^\text{HYB}$ correspond to the exact and hybrid densities. (Lower) The densities for crossings (A) and (C) are benchmarked against the exact, LDA and HF cases; the hybrid, HF and exact curves lie close together.}
\label{figure:3-harmonic-well}
\end{figure}

\textit{Performance for exchange-dominated systems} -- In Fig.~\ref{figure:3-harmonic-well} we demonstrate for the harmonic well with angular frequency $\omega=0.25$ (an exchange-dominated system) that application of any of the conditions (A)--(C) yields an $\alpha$ very close to pure HF, i.e., $\alpha \approx 1$, as expected. Other exchange-dominated systems we tested yield similarly good results from the constrained hybrid.

Conditions (A)--(C) correspond to three `crossing points', as shown in Fig.~\ref{figure:3-harmonic-well}. Using the argument laid out previously, the self-energy should satisfy all three of these conditions. Generally, (A)--(C) correspond to different conditions that specify where the HOMO, LUMO and IP of a system lie with respect to one another. Although it clear from Fig.~\ref{figure:3-harmonic-well} that the three conditions cannot be exactly satisfied, the three crossing points lie pleasingly close together, and the density error $\Delta\rho$ (see Figure caption) is small in their vicinity. Generally, we find that densities obtained from $\alpha$ values lying between crossing points (A) and (C) are in excellent agreement with the exact case. 

\begin{figure}[t!]
\centering
\includegraphics[width=3.3in]{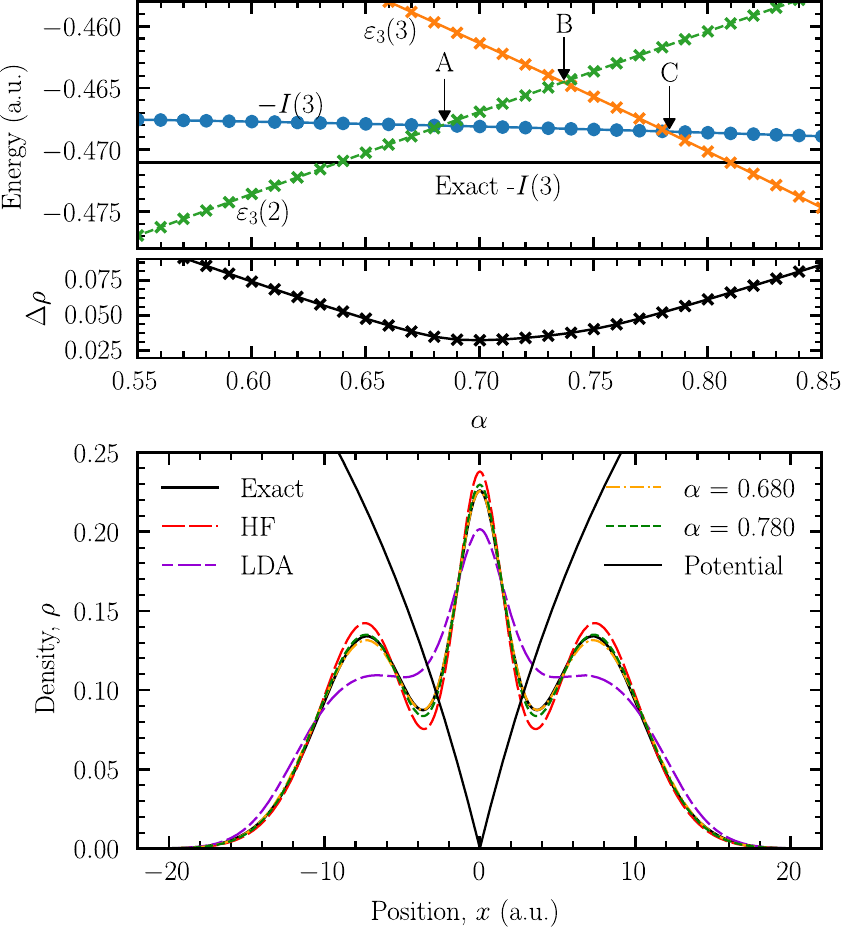}
\caption{ \small As Fig.~\ref{figure:3-harmonic-well}, for three electrons in an atom-like external potential  ($V_\mathrm{ext}\left(x\right)=-1/\left(0.05|x|+1\right)$). The system is correlated as HF fails to predict the exact density and energy.}
\label{figure:3-electron-atom}
\end{figure}

\textit{Performance for correlated systems} -- Given that both of the underlying functionals usually fail to produce a near-exact density in these systems, we ask: is a hybrid functional capable of reproducing a near-exact density for any value of $\alpha$? We show the results in Fig.~\ref{figure:3-electron-atom} for three electrons in an atom-like potential. Once again, all three conditions (A)--(C) produce values of $\alpha$ that yield strikingly accurate densities \footnote{Other correlated systems are shown in the supplemental material}.

Although in the exchange-dominated case crossing points (A) and (C) correspond to an $\alpha$ differing by only one percent, in the correlated system we find that they differ more ($\sim 10\%$). Crucially, however, the density error $\Delta\rho$ corresponding to condition (A) and (C) is better than 0.03. Hence, as before, each density corresponding to these conditions is in excellent agreement with the exact. We note that condition (A) corresponds to a \textit{slightly} better density than (C), the GKT, for both this correlated system and the exchange-dominated system. The alternative hybrid strategy of mixing only the exchange potentials yields accurate, but slightly inferior, densities \footnote{See supplemental material}. 

\begin{figure}[t!]
\centering
\includegraphics[width=3.3in]{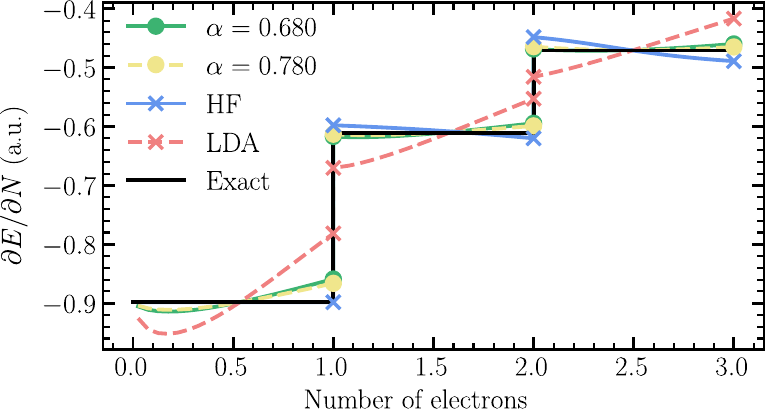}
\caption{ \small The derivative of energy with respect to total number of electrons $N$, $\partial E/\partial N$, for a number of approximations. The external potential and $\alpha$ values chosen are the same as that of Fig.~\ref{figure:3-electron-atom}. We verified that the $\partial E/\partial N$ curve lies exactly on that of the HOMO eigenvalue within each approach. Each node at integer numbers of electrons corresponds to the HOMO and LUMO, with the lower energy value being the HOMO. }
\label{figure:3-electron-atom-dedn}
\end{figure}

In order to verify that the curvature $\partial^2 E/\partial N^2$ in our functionals is indeed better using the constrained hybrid approach, we calculate the derivative of energy with respect to number of electrons $\partial E/\partial N$, shown in Fig.~\ref{figure:3-electron-atom-dedn} \footnote{As calculated in Reference \onlinecite{PhysRevB.18.7165}}.  It can be seen that the HF case is exact for values leading up to one--electron, however curvature is present for anything larger. This is as expected, as the HF energy and density are exact for one electron systems. Unlike HF, the LDA is inexact for all numbers of electrons. The $\alpha$ values corresponding to conditions (A) and (C) in the atom-like potential follow the exact line much more closely than the LDA and HF between 2 and 3 electrons, the region where conditions (A)--(C) have been imposed. This suggests that the curvature has indeed been reduced. Comparing the curvature for conditions (A) and (C), we see that the two are comparable to one another. 

\begin{figure}[t!]
\centering
\includegraphics[width=3.3in]{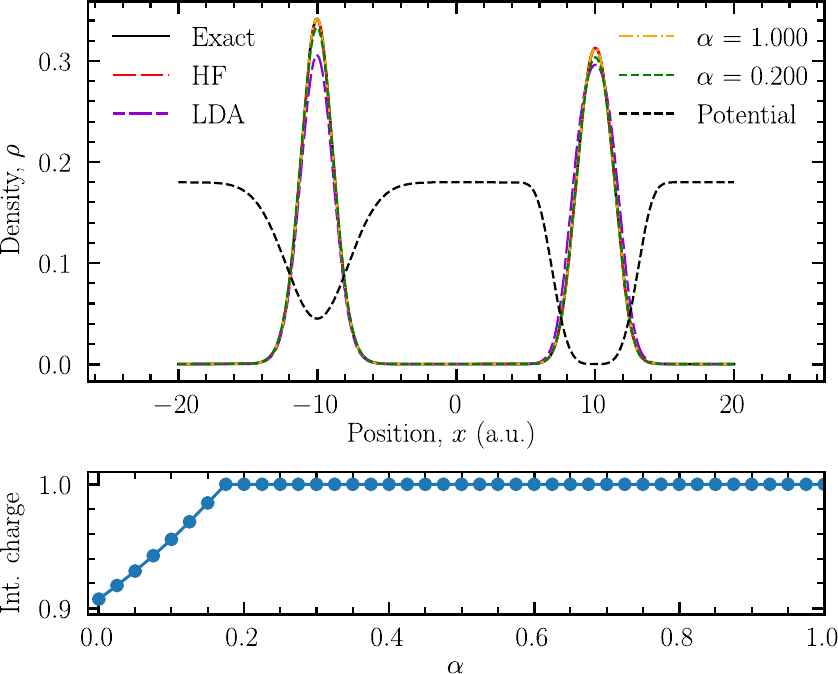}
\caption{ \small (Upper) Densities for various approximations are shown for an exchange-dominated asymmetric double-well potential. The dashed line, illustrating the potential (scaled by 0.15), shows that the two wells are asymmetric. The HF case follows the exact one, placing one electron in each well of a strongly localized system. The LDA predicts that an additional 0.1 electrons are present in the deeper well. The GKT yields $\alpha\approx1$, effectively HF. We show the density for $\alpha=0.2$, which places the correct charge in each well, but has an incorrect density shape. (Lower) The integrated charge of the left (shallower) well is shown for a range of $\alpha$ values.}
\label{figure:dissociation-problem}
\end{figure}

\textit{Fractional dissociation problem} -- We now demonstrate that hybrids are capable of rectifying the fractional charge problem common to many xc approximations for molecular dissociation. Specifically, we test a system with two separated wells where the usual DFT approximations inaccurately predict the amount of charge present in each well. Figure~\ref{figure:dissociation-problem} demonstrates that, when compared with the exact case, the constrained hybrid approach and HF yield near-exact densities. In addition, we show that even for a small fraction of exact exchange ($\alpha=0.200$), the correct charge in each well is obtained, and hence a large range of values of $\alpha$ yield accurate densities. However, the density has an incorrect shape within each well when an $\alpha$ not corresponding to conditions (A)--(C) is used \footnote{We anticipate than bonded open-shell atoms would constrain $\alpha$ more closely.}.

We now show in Table \ref{GAPS} that the accuracy of hybrid functionals for densities is not at the expense of energies. Of particular interest is the quasiparticle energy gap $(I-A)$, which the LDA and HF usually under- and over-estimate, respectively, as well as the values of $I$ and $A$ individually. This establishes contact with the performance of Koopmans-compliant hybrids in 3D systems~\cite{doi:10.1021/ct2009363,PhysRevLett.105.266802,PhysRevB.84.075144} and suggests that useful quasiparticle energies can be extracted from functionals which also produce an accurate density. The tendency of constrained hybrids to reduce these energy gaps from HF to near-exact levels further supports the idea that this approach is similar to a screened-exchange method.

\begin{table}
\caption{The quasiparticle gap of two-electron systems as extracted from the LDA, HF and hybrid\footnote{Constrained using condition (C), though (A) and (B) yield similar results.} HOMO-LUMO eigenvalue differences, compared to the exact gap calculated from many-body total energies. Gaps are compared for the exchange-dominated (harmonic) and correlated (atom-like) systems. The two-electron IPs are shown for the same systems. }
\label{GAPS}
\resizebox{\columnwidth}{!}{
\begin{ruledtabular}
\begin{tabular}{lrrrrr}
(a.u.)		&	LDA		& 	HF		&  Hybrid	& Exact		\\ \hline\hline
\multicolumn{5}{c}{Quasiparticle gaps}						\\ \hline  
Harmonic	&	0.222	&	0.491	& 0.472		& 0.469		\\
\% error	&	53\%	&	5\%		& 1\%		& --	  	\\ \hline
Atom-like	&	0.037	&	0.172	& 0.152 	& 0.141  	\\ 
\% error	&	74\%	&	22\%	& 8\%		& --	 	\\ \hline\hline
\multicolumn{5}{c}{Ionization potentials}					\\ \hline  
Harmonic	&	-0.761	&	-0.620	& -0.629	& -0.628	\\
\% error	&	21.2\%	&	1.3\%	& 0.2\%		& --	  	\\ \hline
Atom-like	&	0.551	&	0.620	& 0.608 	& 0.612  	\\ 
\% error	&	9.9\%	&	1.4\%	& 0.5\%		& --	 	\\
\end{tabular}
\end{ruledtabular}}
\end{table}

\textit{Conclusion} -- Through direct comparison of solutions to the exact many-body Schr\"odinger equation, we have shown that hybrid functionals yield accurate densities {\it and} quasiparticle energy gaps in both exchange-dominated and correlated systems, if the fraction of exact exchange, $\alpha$, is chosen using physically justified constraints, such as the generalized Koopmans' theorem. Particularly accurate densities are obtained from a hybrid strategy that mixes LDA correlation, as well as LDA exchange. The three studied constraints are all in close agreement with one another and all yield accurate densities and gaps. In double-well systems, we find that hybrid functionals perform well and are free from the fractional dissociation problem for a large $\alpha$ range. A key perspective is the interpretation of a hybrid method as a simple screened-exchange approximation within many-body perturbation theory.

\begin{acknowledgments}
K.P.M.\ acknowledges support from EPSRC (EP/K003151/1 and EP/P006051/1) and J.W. acknowledges funding from the York Centre for Quantum Technologies (YCQT). We thank Michele Casula and Mike Entwistle \cite{lda2} for use of the HEG-based LDA and Phil Hasnip for many fruitful discussions.
\end{acknowledgments}
\bibliography{references.bib} 
\end{document}